# On sound-based interpretation of neonatal EEG

S. Gómez, M. O'Sullivan, E. Popovici
Electrical and Electronic Engineering,
University College Cork,
, Ireland
sergi.gomezquintana@ucc.ie

S. Mathieson, G. Boylan, A. Temko
Irish Centre for Fetal and Neonatal Translational Research,
University College Cork, Ireland

*Abstract*—Significant training is required to visually interpret neonatal EEG signals. This study explores alternative sound-based methods for EEG interpretation which are designed to allow for intuitive and quick differentiation between healthy background activity and abnormal activity such as seizures. A novel method based on frequency and amplitude modulation (FM/AM) is presented. The algorithm is tuned to facilitate the audio domain perception of rhythmic activity which is specific to neonatal seizures. The method is compared with the previously developed phase vocoder algorithm for different time compressing factors. A survey is conducted amongst a cohort of non-EEG experts to quantitatively and qualitatively examine the performance of sound-based methods in comparison with the visual interpretation. It is shown that both sonification methods perform similarly well, with a smaller inter-observer variability in comparison with visual. A post-survey analysis of results is performed by examining the sensitivity of the ear to frequency evolution in audio.

I. INTRODUCTION

Neonatal seizures are the most common sign of acute neonatal encephalopathy. Failure to detect such events and the resulting lack of treatment can result in potentially life-threatening outcomes. It is estimated that only 34% of neonatal seizures present clinical signs, the remainder can only be diagnosed using electroencephalography (EEG) monitoring [1]. Previous studies have shown EEG monitoring drastically improves the percentage of correct diagnoses compared to diagnosing seizures based on clinical manifestations alone [1].

Visual interpretation of EEG signals requires significant training and this expertise is often available only in tertiary care units. Even in such centres, it is not available on a 24-hour basis, 7 days a week. In order to ameliorate this situation, a simpler form of EEG called amplitude integrated EEG (aEEG) is often used where 1-2 channels of EEG are recorded and converted to a compressed trend of EEG amplitude over time. However, the use of aEEG amongst the neonatal population has several limitations, waveform information is lost and its effectiveness in seizure detection varies with experience and is poor when compared to EEG [2]. Significant research has been conducted in the area of objective detection of seizure events using artificial intelligence [3-6]. These algorithms aim to provide clinicians with a support in diagnosing abnormal EEG activity, and can achieve accurate seizure detection, though no algorithm will detect all seizures or perform without any false alarms. However, objective methods need to be accompanied with subjective methods such as visualisation of EEG traces to assure that the clinician is engaged in the final decision making.

Methods for EEG sonification aim to facilitate EEG interpretation in a quicker and easier way. In fact, the perception of evolution in frequency over time and the presence of structure (rhythm), which is characteristic of abnormal events such as seizures [7], are more identifiable with hearing rather than with visual aids. Sound-based interpretation aims to mitigate this effect and release the visual sense for other tasks [8]. Several techniques of EEG sonification have been proposed for various applications in the area of adult EEG interpretation, in particular for detection of epileptic discharges [10-13]. Neonatal EEG is different to adult EEG; preterm EEG is different to full term EEG, and even preterm EEG differs dramatically across gestational ages. In previous work, the EEG signal has been usually bottle-necked to just a few core features which are used to drive the sonification process. The auditory representations proposed in this work preserve the structure and completeness of the original EEG signal, consider and analyse the ability to discriminate patterns in the chosen clinical task (intelligibility) and user preferences as a part of the auralisation process. The analysis performed is relevant for the new proposed sonification method as well as the phase vocoder based algorithm which was previously developed for neonatal EEG interpretation [14, 15].

In this paper, the use of frequency and amplitude modulation (FM/AM) in the space of neonatal EEG sonification is explored. FM has been previously used in the space of adult EEG sonification as in [16]. The algorithm implemented in this paper differs in its additional use of AM and customised compression techniques. The proposed algorithm is tuned to increase the sensitivity of the human ear to the presence of rhythm. The phase vocoder algorithm is further explored and its parameters are fine-tuned in order to obtain the highest accuracy of correctly diagnosed neonatal EEG seizure activity. The survey is conducted and the analysis of results is performed along with the subsequent examination of the ear sensitivity to changes in frequency.

II. METHODS AND MATERIALS

*A. Database*

A database of 100 seizure and 100 non-seizure segments was created for the purpose of this study, to be used in the survey.

In order to create this database of 200 segments, a larger database which was previously used for seizure detection algorithm development was utilized [4, 8]. This large database consists of long unedited multichannel EEG recordings from 18 newborns totalling 816 hours of duration with 1389 seizures

Research was supported in part by Wellcome Trust Seed Award (200704/Z/16/Z), SFI INFANT Centre (12/RC/2272), SFI TIDA (17/TIDA/504) and HRB (KEDS-2017-020).

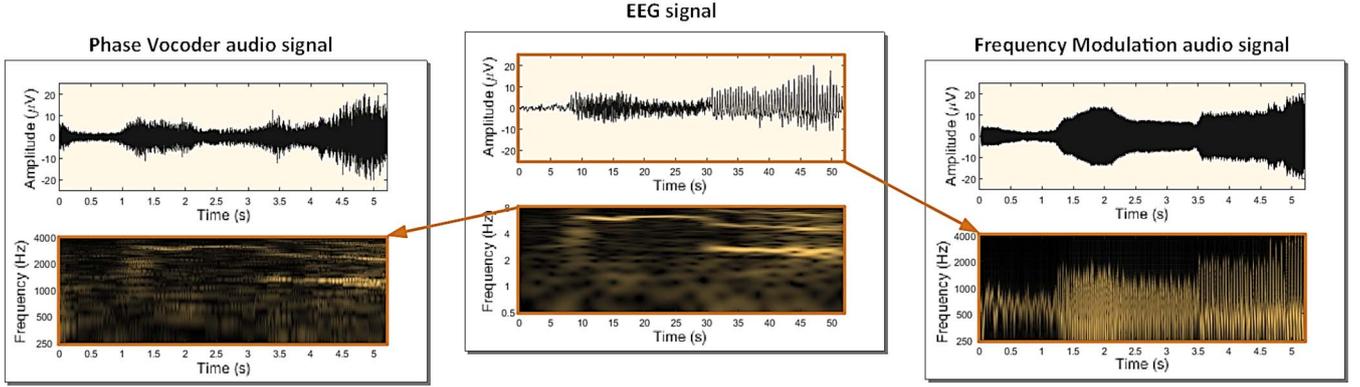

Figure 1. Signal processing chart for PV and FM/AM. PV performs a spectra-to-spectra EEG-to-audio mapping, whereas FM/AM performs a waveform-to-spectra EEG-to-audio mapping. Audio signals are time compressed by a factor of 10 in this example.

annotated by a neonatal neurophysiologist. The dataset contained a wide variety of seizure types including both electrographic-only and electro-clinical seizures of focal, multi-focal and generalized types. Therefore, this dataset is truly representative of the real-life situation in the NICU.

From the 1389 annotated seizures, only a small fraction was annotated on a per-channel basis and rarely the whole seizure from the beginning and the end had a per-channel annotation. The seizure detection algorithm presented in [8] was first used to create AI-based per-channel annotations for all seizures in the dataset. Since the dataset was used to train the algorithm, its performance in a patient-specific setting is very high [9]. On seizure segments with overall seizure annotations, the maximum probability of the seizure detection algorithm was used to select a *single* channel in which the seizure had the highest support of the algorithm. Only a single channel was selected in order to minimize the chances of the error in this process, as seizures can be focal, multi-focal and even migrate across channels.

From a pool of the per channel annotated seizures, the lowest amplitude seizure events of 1-5 min duration were selected from each patient to avoid seizures that were too short or too long, seizures which were easily identifiable by amplitude alone and to preserve the variety of patients, resulting in approximately 200 examples. These examples were further reviewed by an experienced neurophysiologist to discard seizures which were still too obvious or events which were incorrectly selected by the algorithm. This resulted in 100 seizure examples which were confirmed by the clinician.

Using the seizure annotations, from the same dataset of long EEG recordings, the non-seizure segments were extracted to have at least 5 min clearance from the nearest seizure. Subsequently, 200 background EEG segments with the highest amplitude and an algorithmic probability of seizure of at least 0.25 were selected, similarly to avoid any trivial segments, segments which are easily identifiable by amplitude alone, and to preserve equal representation of background activity from each patient. These were then reduced manually to 100 by a neonatal neurophysiologist, by confirming true non-seizure segments and discarding evident high-amplitude artifacts. The resulting 100 non-seizure segments were then randomly cropped in order to have the same distribution of lengths for seizure and non-seizure.

The resulting database was constructed therefore in such a way that seizure identification could not be performed based on amplitude or length and many non-seizure events had seizure-like activity such as respiratory, ECG, sweating or movement artifacts.

### B. Neonatal EEG sonification algorithms

Seizures are partially defined by their slowly decreasing dominant frequencies [17]. This evolution differentiates seizures from artifacts such as respiratory or electrocardiogram [18], which lack temporal frequency evolution.

The Phase Vocoder (PV) algorithm previously developed, allows for the perception of such frequency changes over time in the audio [14, 15]. The PV performs spectra to spectra mapping by preserving a horizontal phase coherence [19] as shown schematically in Fig. 1. The spectra of the EEG signal are preserved in the audio signal obtained with PV.

The FM/AM algorithm presented in this study performs waveform to spectra mapping as illustrated in Fig. 1. The waveform amplitudes of the EEG signal become the spectra of the audio signal. The presence of any structure in the signal (repetitive waveforms) will be perceived as a rhythm in the resultant audio.

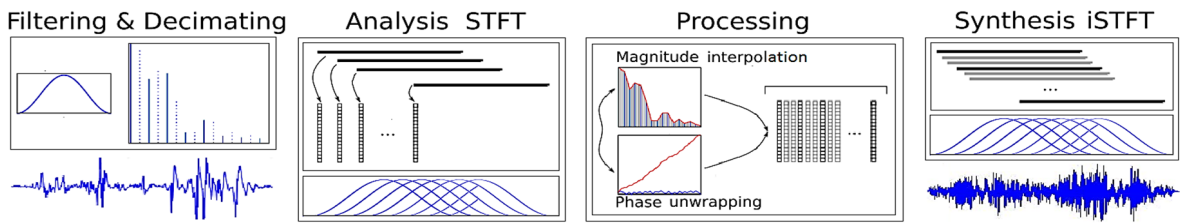

Fig. 2. PV methodology: the EEG signal is processed by frames and reconstructed with coherent phases.

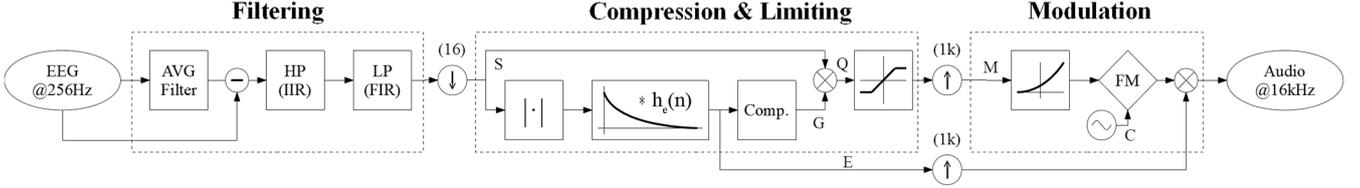

Fig. 3. FM/AM methodology: the EEG signal is filtered, compressed and modulated.

The next subsections provide further details of both algorithms.

*1) Phase vocoder*

PV is an analysis-synthesis method used for scaling frequency while maintaining phase coherence and preserving the spectral envelope of the original EEG signal [18]. The developed PV algorithm converts the low frequency EEG signal (0.5-8Hz) into the audible frequency domain (250-4000Hz). The time length of the resulting output signal can be preserved to match the input signal, or altered to be faster/slower. The effect of time scaling on the discrimination of healthy versus non-healthy EEG is explored in the survey. The algorithm can be broken into four blocks.

*Pre-processing:* EEG signal is band-pass filtered between 0.5-8Hz and down-sampled to 16Hz. These parameters were tuned towards maximising the perception of seizure frequency evolution and are different from those in [15].

*Analysis:* Short Time Fourier Transform (STFT) with a window length/shift of 64s/16s was used. This results in the decomposition of the EEG segment into magnitude and phase for each frequency bin.

*Processing:* The magnitude of the signal is linearly interpolated by a variable time compression factor and saved at an 8kHz sampling rate. The phase is measured and unwrapped to track the cumulative phase variation in order to preserve the phase consistency across interpolated frames.

*Synthesis:* An inverse STFT and overlap-add is applied using the window from the analysis stage, resulting in an 8kHz audio signal.

*2) Frequency & amplitude modulation*

FM and AM synthesis, commonly used in telecommunication, is the process of varying a carrier wave frequency and amplitude proportionally to a modulating wave. In this work, the EEG is used as the modulating waveform. The developed algorithm incorporates a number of digital signal processing techniques as outlined in Fig. 3.

*DC Removal:* DC removal is achieved by applying a moving average filter to the input signal. As the lowest frequency of interest is 0.5Hz, a minimum window length of 2 s is required.

*Pre-processing:* EEG signal is band-pass filtered between 0.5-7.5 Hz and down-sampled to 16Hz.

*Compressor:* EEG signals often contain amplitude spikes that exceed the common amplitude range of ±50μV. In order to avoid aliasing in the FM stage, compression is applied to reduce the dynamic range of the input signal ($S$) with minimal distortion. For that purpose, the envelope ($E$) is calculated as the convolution on the absolute value of $S$ using an exponential impulse response $h_e(n)$ with a decay time of 8s. Compressors operate on a logarithmic scale, in decibels (dB). Thus, the input signal ($E$) is firstly converted to ($E_{dB}$) using Eq. 1:

$$E_{dB} = 20 \times \log|E| \qquad (1)$$

The envelope ($E_{dB}$) is used to control the gain reduction. When the amplitude of the envelope is under the specified threshold ($T_{dB}$), no gain reduction is applied to the signal. However, when the threshold is exceeded, the amplitude of the signal is reduced proportional to the compression ratio ($R$) as in Eq. 2:

$$G_{dB} = \begin{cases} 0 & E_{dB} < T_{dB} \\ -\left(1 - \frac{1}{R}\right)(E_{dB} - T_{dB}) & otherwise \end{cases} \qquad (2)$$

For every 1 dB that the input signal exceeds $T_{dB}$, the amplitude of the output will be reduced by an amount of $(1 - 1/R)$. As $R \geq 1$, the gain reduction ($G_{dB}$) is always $\leq 0$. $G_{dB}$ is subsequently converted back to the linear domain ($G$) by using the reverse of Eq. 1. $G$ is then used to control the gain of the input $S$ to obtain the output $Q$ as seen in Fig. 3.

In this algorithm, $T_{dB}$ of -20dB (5μV) was chosen, which is 1/10 of the expected full-scale amplitude. $R$ was chosen to be 1.5, meaning that a spike of 150μV would be reduced to 50μV, keeping the output within the desired dynamic range.

The output, $Q$, is then amplified from ±50μV to ±1V, and hard-limited as in Eq. 3 to ensure the signal is free from aliasing when used as the modulator signal in the subsequent FM stage:

$$M(Q) = \begin{cases} 1 & Q > 50\mu V \\ 2 \times 10^4 Q & -50\mu V < Q < 50\mu V \\ -1 & Q < -50\mu V \end{cases} \qquad (3)$$

*FM:* The modulator signal ($M$) is firstly up-sampled (x 1000) to 16kHz in order to achieve an audio bandwidth. As the human hearing system perceives frequency on an exponential scale [20], an exponential transform is applied to $M$. If $M$ is 0V, then the value of the frequency ($F$) is 500Hz, if $M$ is -1V, then $F$ goes down to 50Hz and if $M$ is 1V, $F$ goes up to 5kHz as in Eq. 4:

$$F(M) = 500 \times 10^M \qquad (4)$$

The resulting output wave, $W$, is a frequency-varying signal [21], in which the amplitude variations of the EEG are mapped to the frequency variations of the output signal, based on an sine-wave FM synthesiser.

*AM*: The envelope $E$ is used to modulate the amplitude of $W$. Thus, the amplitude of the EEG signal is also embedded in the output audio.

### 3) Time compression

Both sonification algorithms can be used to manipulate the time scale, so that the duration of the audio output can be different to that of the input EEG. This is convenient for generating faster sonification, allowing for the review of long EEG signals in much shorter segments of audio. This characteristic is also investigated for the discrimination and perception of changes in the EEG morphology, which often evolve too slowly to be perceived by the human ear in real time.

The time compression factor ($r \geq 1$) defines the relationship between the durations of the EEG and the resulting audio. For instance, to scale 60 mins of EEG into 6 mins of audio, a value of $r = 10$ will be used, $r = 1$ results in direct 1:1 time scaling.

In PV, time compression is applied in the processing stage when interpolating the magnitude and phase. The interpolation factor is decreased with respect to $r$. In FM/AM, the time compression is applied by simply reducing the value of the up-sampler by $r$.

## III. RESULTS & DISCUSSION

### A. Survey design

In order to quantitatively assess the clinical performance of the sonification algorithms, a survey was conducted amongst a small cohort ($N = 11$) of non-EEG-expert participants. The survey tests and compares the seizure identification rate using seven tested scenarios – visual means (scenario 1), PV with a time-compression factor of 1, 5, and 10 denoted by PVx1, PVx5, and PVx10, scenarios 2 to 4; and FM/AM with a time-compression factor of 1, 5 and 10 denoted by FMx1, FMx5 and FMx10, respectively, scenarios 5 to 7.

Training data (3 seizures and 3 non-seizure segments) were presented to each survey participant. The data were processed according to the tested scenario for each method. During the survey, 10 examples of neonatal EEG were randomly selected from the created database for each scenario. The participant was required to mark each example as seizure or non-seizure. Each participant sequentially went through each of 7 tested scenarios. The individual percentage scores of correct diagnoses were calculated.

### B. Sonification survey results

Fig. 4 shows the results of the survey as the mean and 95% confidence intervals. It can be seen that both FM/AM and PV sonification methods when speeded up by a factor of 10 perform equally well, at 76% and 73% accuracy, respectively. These methods slightly outperform visual interpretation, which obtains 69%. PV at a speed of 1, achieves the best real-time (1:1) sonification result at 69%. The variance of visual interpretation is the largest denoting its inconsistent accuracy across participants.

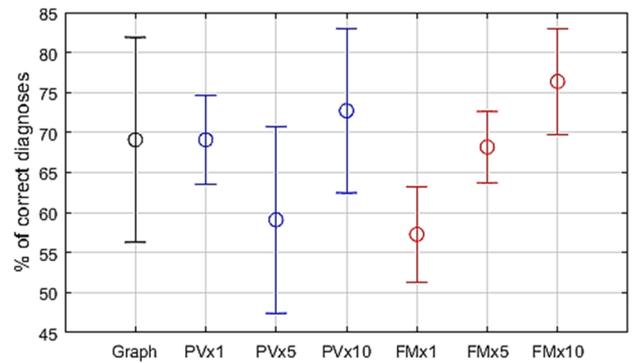

Figure 4. Sonification survey results.

Fig. 5 shows the same results divided into 2 groups. Group A ($N = 3$) includes the participants who performed below the average correct rate for visual interpretation, and group B includes the remainder of the cohort (i.e. above the average, $N = 8$). It can be seen that visual interpretation differs by more than 40% between the two groups. For sonification methods, the difference is smaller than 15 % in the worst case. Moreover, the same trends in the sonification methods are preserved across both groups, with the increased speed generally resulting in the increased performance of correct diagnoses. These results show that audio interpretation is more consistent.

The participants were also asked about their preferred method of EEG sonification. Three participants preferred PV and eight preferred FM. Among them, two gave preference to PV×10 and four chose FM×10. As the question was answered without knowing the accuracy achieved with each method, this result represents the aesthetic choice of the sonification which is complementary to the accuracy-based assessment in Fig. 4.

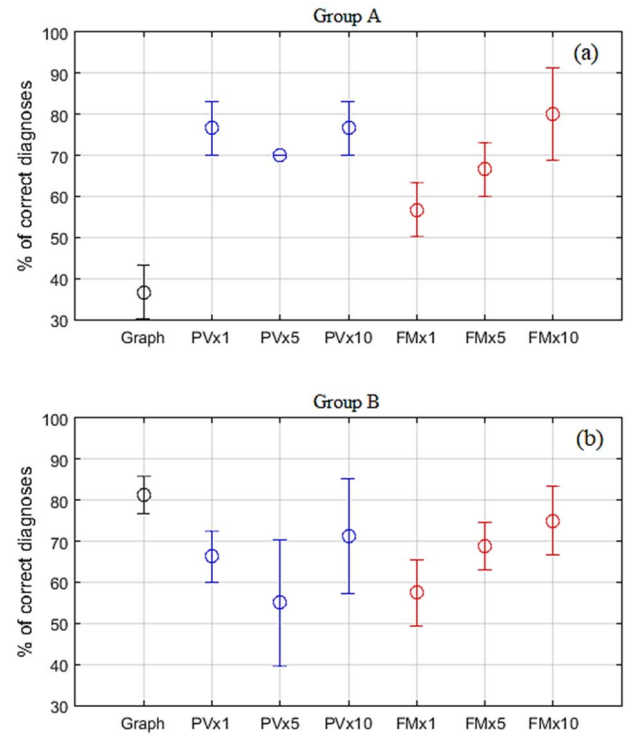

Figure 5. Sonification survey results by groups. Plot (a) shows the group with an accuracy below average on visual assesment. Plot (b) shows the group with the accuracy above average.

*C. Post-survey analysis of seizure frequency evolution*

Previous studies have shown that during seizures, the dominant frequencies in the EEG evolves over time, decreasing in many cases [17]. However, that evolution can be so slow that for some sonification methods the human hearing sense is incapable of distinguishing a change in the frequency of the tone. In order to generate a comparison between the evolution of the seizure frequency and human hearing sensitivity of tone change, a model of seizure frequency evolution and human sensitivity of frequency change is required. A suitable figure of merit for change in frequency is octaves per minute (oct/min), which is the logarithmic measure of change in frequency over time. A change of +1 oct/min or -1/oct/min implies that in 1 minute the frequency doubles or is divided by 2, respectively. The relative change of frequency over time can then be expressed as:

$$\Delta f [oct] = \log_2 \frac{f_2}{f_1} \quad (5)$$

$$slope \left[\frac{oct}{min}\right] = \frac{\Delta f [oct]}{\Delta t [min]} \quad (6)$$

where $\Delta f$=relative change in frequency in octaves, $f_1$, $f_2$= start & end frequencies, $slope$= change in frequency over time.

The seizure and non-seizure examples were analysed with respect to the evolution of the dominant frequencies in each EEG epoch. Each EEG example is segmented into 2s epochs with 1 sample shift and the STFT is computed. The location of the dominant frequency is found. Since the time difference between epoch is 1 sample, the number of octaves between two consecutive epochs is the slope in oct/sample, that can be converted to oct/min as shown in Eq. 7:

$$slope \left[\frac{oct}{min}\right] = slope \left[\frac{oct}{sample}\right] \times \frac{16 samples}{1s} \times \frac{60s}{1min} \quad (7)$$

Fig. 6 shows the distribution of frequency evolution rates for the seizure and non-seizure examples in the database. It can be seen that the peak of the distribution of the frequency changes for seizures is at -0.373oct/min. This indicates that there is a negative slope associated with the seizure events, denoting a decrease in frequency over time. For example, if the dominant frequency of the seizure was 6Hz, then after 60s, on average, it will decrease to 4.6Hz. In comparison, the peak of the distribution for non-seizure is at -0.01oct/min, which implies that there are no consistent changes in the frequency for background EEG and seizure-like artifacts.

From Fig. 6 it can also be seen that in the range between -1.69 oct/min and -0.13 oct/min, the seizure slope distribution is above the non-seizure slope distribution and seizures are separable from non-seizures. This indicates that the measure can also be of use in automated seizure detection algorithms.

*D. Human sensitivity to frequency evolution*

A test of human sensitivity to changes of continuous frequency was performed to gain further insight in the results from Fig. 4. A small survey was conducted with a cohort of 6 participants who repeated the test multiple times. In this survey, the participants were asked to listen to a tone. The frequency of the tone was randomly chosen in the range between 500Hz and 3500Hz. The frequency linearly varied over time, with a randomly generated slope in the range between -0.5oct/min and 0.5oct/min. Participants were asked a binary question of whether they perceived increase or decrease of a frequency. The results were compiled for every slope. The power density function (PDF) can be constructed by computing the ratio between errors and trials.

The distribution of human errors in the sensitivity survey is shown in Fig. 6. The distribution is centred around 0 and 95% of human errors lie in the range between -0.389oct/min and 0.285oct/min. These results explain why speeding up the audio resulted in a better discrimination between seizure and non-seizure. For a factor of 10, the slope of the frequency evolution for seizures is pushed outside the range of slopes in which the human hearing system is incapable of distinguishing a change in frequency. The perception of the evolution of the low-frequency activity is facilitated with time compression.

IV. CONCLUSION

A non-expert EEG interpretation survey is presented and provide promising results for the application of sonification in neonatal seizure detection. Both methods of sonification performed similarly well. Visual interpretation resulted in the largest variance, denoting that non-expert visual detection of seizures is inconsistent. It is observed that speeding up the EEG in audio results in a higher accuracy. Analysing the frequency evolution of seizures, it was shown that the peak value of the frequency slope is -0.373oct/min. The human sensitivity to changes in continuous frequency indicated a decreased sensitivity to changes in frequency with a slope inside the range of -0.389 to 0.285 oct/min. Thus, sonification methods with larger time-scale factors will result in more accurate sound-based seizure identification.

Future work will expand the results to include a larger cohort and targeting clinical end-users. The parameters of sonification algorithm will be tuned according to the objective results obtained in the frequency evolution and human sensitivity to frequency tests. The optimised sonification algorithm will be implemented in the portable acquisition and interpretation system presented [21], providing a viable medium to bring sonification into clinical practice.

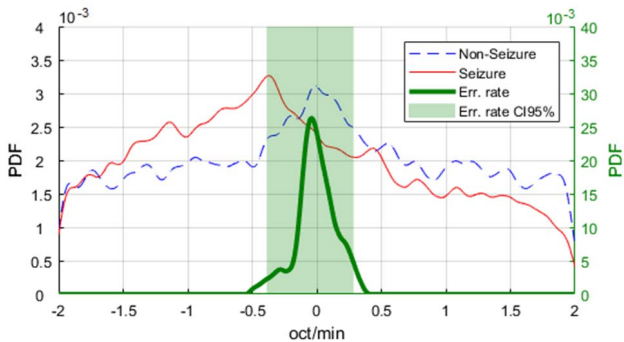

Figure 6. Estimated PDF of seizure and non-seizure frequency slope distribution (red and blue respectively). Green area depicts the 95% confidence interval where human hear are less accurate at detecting frequency change. Best viewed in color.

## REFERENCES

[1]. D. Murray, G. Boylan, I. Ali, C. Ryan, B. Murphy and S. Connolly, "Defining the gap between electrographic seizure burden, clinical expression and staff recognition of neonatal seizures," *Archives of Disease in Chidhood: Fetal and Neonatal Edition,* vol. 93, no. 3, pp. 187-191, 2007.

[2]. A. Rakshasbhuvankar, S. Paul, L. Nagarajan, et al., "Amplitude-integrated EEG for detection of neonatal seizures: a systematic review," *Seizure*, v. 33, pp. 90-98, 2015.

[3]. S. Mathieson, N. Stevenson, E. Low, W. Marnane, J. Rennie, A. Temko, G. Lightbody and G. Boylan, "Validation of an automated seizure detection algorithm for term neonates," *Clinical Neurophysiology*, vol. 127, no. 1, pp. 156-168, 2016.

[4]. E. Thomas, A. Temko, W. Marnane, G. Boylan, G. Lightbody, "Discriminative and generative classification techniques applied to automated neonatal seizure detection," *IEEE Journal of Biomedical and Health Informatics*, v. 17, i. 2, pp. 297-304, 2013.

[5]. N. Stevenson, I. Korotchikova, A. Temko, G. Lightbody, W. Marnane and G. Boylan, "An Automated System for Grading EEG Abnormality in Term Neonates with Hypoxic-Ischaemic Encephalopathy," *Annals of Biomedical Engineering*, vol. 41, no. 4, pp. 775-785, 2013.

[6]. AH. Ansari, P. Cherian, A. Caicedo Dorado, G. Naulaers, M. De Vos, S. Van Huffel, "Neonatal Seizure Detection Using Deep Convolutional Neural Networks," *International Journal of Neural Systems*, 2018.

[7]. S. Patrizi, G. Holmes, M. Orzalesi and F. Allemand, "Neonatal seizures: characteristics of EEG ictal activity in preterm and fullterm infants," *Brain and Development*, vol. 25, no. 6, pp. 427-437, 2003.

[8]. A. Temko, W. Marnane, G. Boylan and G. Lightbody, "Clinical implementation of a neonatal seizure detection algorithm," *Decision Support Systems*, vol. 70, pp. 86-96, 2015.

[9]. A. Temko, A. K. Sarkar, G. B. Boylan, S. Mathieson, W. P. Marnane and G. Lightbody, "Toward a Personalized Real-Time Diagnosis in Neonatal Seizure Detection," in *IEEE Journal of Translational Engineering in Health and Medicine*, vol. 5, pp. 1-14, 2017.

[10]. T. Väljamäe, T. Steffert, S. Holland, X. Marimon, R. Benitez, S. Mealla, A. Oliveira and S. Jordà, "A review of real-time eeg sonification research," in *International Conference on Auditory Display (ICAD)*, Łódź, 2013.

[11]. H. Khamis, A. Mohamed, S. Simpson and A. McEwan, "Detection of temporal lobe seizures and identification of lateralisation from audified EEG," *Clinical Neurophysiology*, vol. 123, no. 9, pp. 1714-1720, 2012.

[12]. Jeng-Wei Lin, Wei Chen, Chia-Ping Shen, Ming-Jang Chiu, Yi-Hui Kao, Feipei Lai, Qibin Zhao, Andrzej Cichocki, "Visualization and sonification of long-term epilepsy electroencephalogram monitoring," *Journal of Medical and Biological Engineering*, pp 1–10. 2018.

[13]. T. Hermann, A. Hunt, J. Neuhoff (editors). *The Sonification Handbook*. Logos Publishing House, Berlin, Germany, 2011.

[14]. A. Temko, G. Lightbody, L. Marnane and G. Boylan, "Real time audification of neonatal electroencephalogram (EEG) signals". US Patent US20170172523A1, 07 07 2014.

[15]. A. Temko, W. Marnane, G. Boylan, J. O'Toole and G. Lightbody, "Neonatal EEG audification for seizure detection," in *IEEE International Conference of Engineering in Medicine and Biology Society (EMBC)*, Chicago, 2014.

[16]. J. Parvizi, K. Gururangan, B. Razavi, C. Chafe, "Detectig silent seizures by their sound," *Epilepsia In Press*, vol. 59, no. 4, pp. 877-884, 2018

[17]. A. Krystal, R. Prado and M. West, "New methods of time series analysis of non-stationary EEG data:," *Clinical Neurophysiology*, vol. 110, no. 12, pp. 2197-2206, 1999.

[18]. C. Hagmann, N. Robertson and D. Azzopardi, "Artifacts on Electroencephalograms May Influence the Amplitude-Integrated EEG Classification: A Qualitative Analysis in Neonatal Encephalopathy," *Pediatrics*, vol. 118, no. 6, pp. 2552-2554, 2006.

[19]. J. Flanagan and R. Golden, "Phase Vocoder," in *IEEE The Bell System Technical Journal, Minneapolis*, 1966.

[20]. J. McDermott, M. Keebler, C. Micheyl and A. Oxenham, "Musical intervals and relative pitch: Frequency resolution, not interval resolution, is special," *The Journal of the Acoustical Society of America*, vol. 128, no. 4, p. 1943, 2010.

[21]. J. Chowning, "The synthesis of complex audio spectra by means of frequency modulation," *Computer Music Journal*, vol. 1, no. 2, pp. 46-54, 1977.

[22]. M. O'Sullivan, S. Gomez, A. O'Shea, E. Salgado, K. Huillca, S. Mathieson, G. Boylan, E. Popovici, A. Temko, "Neonatal EEG interpretation and decision support framwork for mobile platforms", in *IEEE International Conference of Engineering in Medicine and Biology Society (EMBC)*, Honolulu, 2018.